\documentclass{iau}
\usepackage{graphicx}

\title[Multicolour photometry of pulsating stars] 
{Multicolour photometry of pulsating stars in the Galactic Bulge fields}

\author[P. Bru{\'s} \& Z. Ko{\l}aczkowski]   
{Przemys{\l}aw Bru{\'s}
 \and Zbigniew Ko{\l}aczkowski}

\affiliation{Instytut Astronomiczny Uniwersytetu Wroc{\l}awskiego \\
Kopernika 11, 51-622 Wroc{\l}aw, Poland \\ email: {\tt brus@astro.uni.wroc.pl} \\[\affilskip]}

\pubyear{2014}
\volume{301}  

\pagerange{1--2}
\jname{Precision Asteroseismology}
\editors{J.A. Guzik, W.J. Chaplin, G. Handler \& A. Pigulski, eds.}
\begin{document}
\maketitle

\begin{abstract}
We present a study of photometric properties of very crowded stellar fields toward the Galactic Bulge. We performed a search for pulsating stars
among thousands of variable stars from the OGLE-II survey supplementing the variability study with
photometric measurements in four Johnson-Cousins $UBVI_{\rm C}$ passbands. Using these data, 
we analysed the properties of objects located at different distances and, whenever possible, classified them.
\end{abstract}

\firstsection 

\section{Archival data and follow-up observations}
The main source of data used in this work is the entire OGLE-II DIA time-series $I_{\rm C}$ band photometric database 
of Galactic Bulge fields (\cite[Szyma{\'n}ski 2005]{Szym2005}). 
All pulsating stars described here were discovered during 
our multiperiodicity search in the OGLE-II database. Most of them are not present in the catalog published 
by the OGLE Team (\cite[Wo{\'z}niak at al.~2002]{Wozniak2002}) because of very small amplitudes (a few mmag).  

The second part of our project is based on follow-up single-epoch observations of 25 selected fields carried 
out with the CTIO 1-m telescope in May and June 2007. Using the Y4kCam detector ($20' \times 20'$ field of view), we covered a relatively 
small part of the OGLE-II fields in four passbands: $U$ (exposures: 1500 s), $B$ (400 s), $V$ (250 s), $I_{\rm C}$ (120 s). 
In each field we performed profile photometry by means of the DAOPHOT-II package (\cite[Stetson 1987]{Stet87}).  
To perform the transformation of our photometry to the standard system, we carried out additional 
observations of a nearby standard field BWC (\cite[Paczy{\'n}ski et al.~1999]{Paczynski1999}) on a 
photometric night of June 3/4, 2007. Accurate astrometry carried out by means of the UCAC3 catalog 
(\cite[Zacharias et al.~2010]{Zacharias2010}) allowed us to perform reliable cross-identifications 
with the OGLE databases. Our standardized $V$ and $I_{\rm C}$ measurements are in very good agreement 
with the OGLE-III photometry published by \cite[Szyma{\'n}ski et al.~(2011)]{Szym2011}.

\section{Analysis and results}
In order to remove long-term and seasonal trends, we applied spline function fits as the first step of the time-series analysis. 
Next, we performed frequency analysis by means of the Fourier periodogram in the range between 0 and 40~d$^{-1}$. 
A given star was selected as a variable if the signal-to-noise ratio (S/N) of the dominant frequency exceeded a detection level equal to 5.  
For all stars selected through the S/N criterion, we applied a semi-automatic search for additional frequencies by means of consecutive 
prewhitening and repeated Fourier analysis of the residual light curve. In each iteration all parameters of the fit were updated. As a result, we found several hundreds of multiperiodic variables in each OGLE-II field. They are  
good candidates for pulsators, mainly those on the main-sequence. Moreover, we divided these stars into two groups: short-period pulsators with
\begin{figure}[t]
 \vspace*{0 cm}
\begin{center}
\includegraphics[angle=-90,width=3.5in]{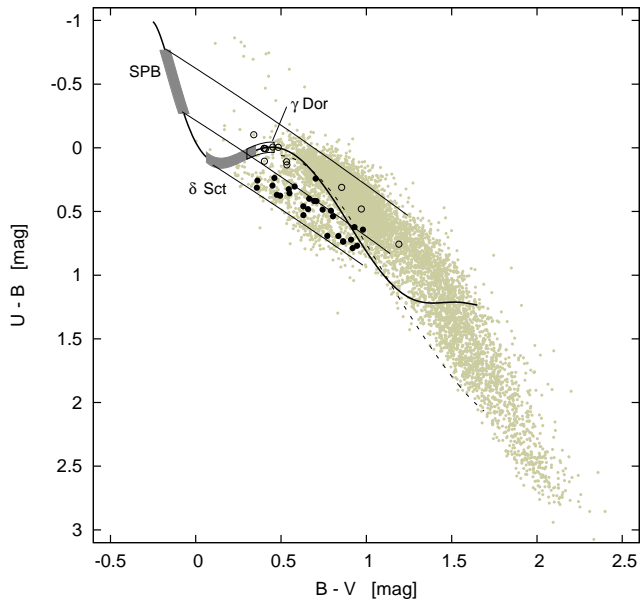}
 \vspace*{-0.1 cm}
 \caption{Observed ($U-B$) vs.~($B-V$) colour-colour diagram for stars in three selected fields toward the Galactic Bulge. 
All stars with ($U-B$) errors smaller than 0.05 mag are shown as gray dots.
The intrinsic relations for dwarfs and giants are plotted as solid and dashed curves, respectively.
The schematic locations of SPB, $\delta$~Sct and $\gamma$~Dor type pulsators are marked and labeled.
Stars classified as $\delta$~Sct-type variables are shown as black filled dots, SPB or $\gamma$~Dor 
variables as open circles. Three solid black lines indicate reddening line with a slope of 0.85.}
\label{fig1}
\end{center}
\end{figure}
dominant periods shorter than 0.3 d and long-period pulsators including the remaining multiperiodic variables. 
In the first group we expect $\beta$~Cephei and $\delta$~Scuti-type stars; the second group consists 
of slowly pulsating B-type stars (SPB) and $\gamma$~Doradus-type stars.

The preliminary multi-colour photometry of three selected fields allowed us to make use of 
colour-magnitude and colour-colour diagrams in our classification of variable stars. 
We found pulsating stars in a wide range of colour indices measured with an accuracy better than 0.05 mag (Fig.~1). 
The distribution of early A-type stars in the ($U-B$) vs.~($B-V$) 
diagram implies a significantly higher slope of the reddening line than the standard value.
The whole sample of short-period variables turned out to contain only $\delta$~Scuti-type stars with $E(B-V)$ reddenings
between 0.3 and 1 mag. 
The majority of the long-period pulsators cannot be unambiguously classified: they can be either highly reddened SPB or $\gamma$~Doradus stars. 

 {\bf Acknowledgments.} This work was supported by the National Science Center (NCN) grant 
No.~2011/03/B/ST9/02667.

\end{document}